\documentstyle[preprint,aps,rotating]{revtex}
\begin{document}
\tightenlines
\input epsf

\setcounter{page}{0}
\thispagestyle{empty}
\begin{center}
\begin{flushright}
\rm \normalsize {CERN-TH/99-175\\
hep-ph/9906415}
\end{flushright}
\vspace*{1cm}
{\large \bf Heavy quark $\bbox{1/m_Q}$ contributions in 
semileptonic $\bbox{B}$~decays 
to orbitally  excited $\bbox{D}$ mesons  }

\vspace*{1cm}
D. Ebert$^{1,2}$, R. N. Faustov$^2$\footnote{On leave of absence 
from the Russian Academy of Sciences,
Scientific Council for Cybernetics,
Vavilov Street 40, Moscow 117333, Russia.}
~and V. O. Galkin$^3$

{ \small\it $^1$Theory Division, CERN, CH-1211 Geneva 23, Switzerland\\
$^2$Institut f\"ur Physik, Humboldt--Universit\"at zu Berlin,
Invalidenstr.110, D-10115 Berlin, Germany\\
$^3$Russian Academy of Sciences, Scientific Council for
Cybernetics, Vavilov Street 40, Moscow 117333, Russia}
\end{center}
\begin{abstract}
Exclusive semileptonic decays of $B$ mesons to orbitally excited $D$ mesons are 
considered beyond the infinitely heavy quark limit in the framework of the relativistic
quark model based on the quasipotential approach. This model agrees
with the structure of heavy quark mass corrections predicted by the heavy quark
effective theory and allows the determination of corresponding leading and subleading
Isgur--Wise functions.  It is found that both relativistic and $1/m_Q$ contributions
significantly
influence the decay rates. Thus, relativistic transformations of the meson wave functions
(Wigner rotation of the light quark spin) already contribute at leading order of the
heavy quark expansion and result in a suppression of $B\to D^{*}_0e\nu$ and
$B\to D^{*}_1e\nu$ decay rates. On the other hand, the vanishing of the decay matrix
elements at zero recoil of a final $D^{**}$ meson in the infinitely heavy quark mass limit 
makes the $1/m_Q$ corrections to be very important, and their account 
results in a substantial enhancement of $B\to D_1e\nu$ and $B\to D^{*}_0e\nu$ decay
rates. 
\end{abstract}
{\vskip 3.5cm
\begin{flushleft} CERN-TH/99-175 \\
June 1999
\end{flushleft}}
\vfill\eject

\setcounter{footnote}{0}
\renewcommand{\thefootnote}{\arabic{footnote}}
\setcounter{page}{1}
\pagestyle{plain}

\section{Introduction}
The investigation of semileptonic decays of $B$ mesons to excited
$D$ meson states is an important problem for heavy flavour physics.
In particular, these decays can 
provide an additional source of information for the determination of the 
Cabibbo--Kobayashi--Maskawa matrix element $V_{cb}$ as well as on the 
relativistic quark dynamics inside heavy--light mesons. The experimental
data on these decays are becoming available now \cite{cleo,aleph,opal},
and the $B$ factories
will provide more accurate and comprehensive data. It is necessary to note that the
experiment shows that only approximately 60\% of the inclusive semileptonic $B$ decay
rate is due to the decays to ground state pseudoscalar $D$ and vector $D^*$ mesons.
Thus the rest of these decays should go to excited $D$ meson and continuum states. 
 
The presence of the heavy 
quark in the initial and final meson states in these decays considerably
simplifies their theoretical description. A good starting point for this
analysis is the infinitely heavy quark limit, $m_Q\to\infty$ \cite{iw}. 
In this limit the heavy quark symmetry arises, which strongly
reduces the number of independent weak form factors \cite{iw1}.
The heavy quark mass and spin then decouple and all meson properties are
determined by light-quark degrees of freedom alone. As a result the heavy
quark degeneracy of energy levels emerges. The spin $s_q$ of 
the light quark couples with
its orbital momentum $l$ ($j=l\pm s_q$), resulting for $P$-wave mesons in 
two degenerate
$j=3/2$ states ($J^P=1^{+},2^+$)\footnote{Here $J=j\pm 1/2$  
is the total angular momentum, and the superscript $P$ denotes
the meson parity.} and two degenerate $j=1/2$ states 
($0^+,1^+$). The heavy quark symmetry also predicts that the 
form factors for $B\to D^{**}e\nu$ decays, where $D^{**}$ is a generic 
$P$-wave $D$ meson state\footnote{For concrete $P$-wave meson states
the standard notations $D_1$, $D_2^{*}$, $D_1^{*}$ and $D_0^{*}$ are used.}, can be expressed in terms of two independent
Isgur--Wise functions  \cite{iw1}. However, in the infinitely heavy quark limit the decay
matrix elements between a $B$ meson and an orbitally excited $D$ meson vanish
at zero recoil because of the heavy quark spin-flavour symmetry \cite{iw1}.
The kinematically allowed range for these decays is not broad. As a 
result the role of relativistic and finite heavy quark mass 
contributions  not vanishing at zero recoil is considerably
more important here than in the decays to ground state $D$ mesons. 
Thus, the magnitude of such corrections might be comparable with the leading order result.   

Recently the first order $1/m_Q$ corrections to the exclusive semileptonic $B$
decays into excited charmed mesons were investigated 
 within the heavy quark effective theory (HQET) \cite{llsw}.
The structure of the $1/m_Q$ corrections to decay matrix elements, which follows from 
QCD and heavy quark symmetry, was determined.  It was found that at the
first order of heavy quark expansion the $B\to D_1$, $B\to D^*_0$ and $B\to D^*_1$
matrix elements do not vanish at zero recoil and can be expressed at this kinematical
point in terms of the leading Isgur--Wise functions. 
Away from the zero recoil point new subleading
Isgur--Wise functions arise, which cannot be determined from symmetry
considerations alone. Thus for their determination, additional model dependent 
assumptions are necessary. In Ref.~\cite{llsw} an estimation of these
functions, based on the non-relativistic quark model as well as on some additional 
assumptions, was  made. It is just here that we can apply the relativistic quark model to get a more
consistent calculation of relevant Isgur--Wise functions and decay rates. 

Our relativistic quark model is based on the quasipotential approach in quantum field
theory with a specific choice of the quark--antiquark interaction potential. It provides
a consistent scheme for the calculation of all relativistic corrections at a given $v^2/c^2$ 
order and allows for the heavy quark $1/m_Q$ expansion. In preceding papers 
we applied this model to the calculation of the mass spectra of 
orbitally and radially excited states of heavy--light mesons \cite{egf},
as well as to a description of weak decays of $B$ mesons to ground state
heavy and light mesons \cite{fgm,efg}. The heavy quark expansion for the ground
state heavy-to-heavy semileptonic transitions \cite{fg} has been found to be 
in agreement with model-independent predictions of the HQET. We considered 
the exclusive semileptonic decays of $B$ mesons to orbitally excited $D$ mesons in the infinitely heavy quark
limit in \cite{efg1} and found the important relativistic contribution to the leading
Isgur--Wise functions arising from the relativistic transformation of the meson wave 
function.

The paper is organized as follows. In Sec.~II we briefly present the necessary HQET
results on the $B\to D^{**}$ transition matrix elements obtained in Ref.~\cite{llsw}.
In Sec.~III we describe our relativistic quark model, putting special 
emphasis on
the calculation of decay matrix elements and on the relativistic transformation of a
meson  wave function from the rest reference frame to the moving one.  The heavy quark 
expansion for decay matrix elements is carried out up to
the first order $1/m_Q$ corrections
and compared to model independent HQET predictions in Secs.~IV-VI.
There we present our
results for leading and subleading Isgur--Wise functions and compare  predictions
for decay rates with and without the $1/m_Q$ corrections being taken into account.
Taking account of $1/m_Q$
corrections leads to a substantial enhancement of $B\to D_1e\nu$ and 
$B\to D^{*}_0e\nu$ decay rates and gives better agreement between theoretical predictions
and available experimental data. We also present the electron spectra for the
considered decays and test the fulfilment of the Bjorken sum rule. 
Section~VII contains our conclusions. 
 
\section{HQET results for decay matrix elements}

\subsection{ $B\to D_1 e\nu$ and $B\to D_2^*e\nu$ decays}

The matrix elements of the vector and axial-vector currents 
 between $B$ mesons and $D_1$ or $D_2^*$ mesons can be parametrized in
the following way
\begin{eqnarray}\label{ff1}
{\langle D_1(v',\epsilon)| \bar c\gamma^\mu b |B(v)\rangle \over \sqrt{m_{D_1}m_B}}
  &=& f_{V_1}\epsilon^{*\mu} 
  + (f_{V_2} v^\mu + f_{V_3} v'^\mu) (\epsilon^*\cdot v) , \cr \cr
{\langle D_1(v',\epsilon)| \bar c\gamma^\mu\gamma_5 b |B(v)\rangle \over \sqrt{m_{D_1}m_B}}
  &=& i f_A \varepsilon^{\mu\alpha\beta\gamma} 
  \epsilon^*_\alpha v_\beta v'_\gamma , \cr \cr
{\langle D^*_2(v',\epsilon)| \bar c\gamma^\mu\gamma_5 b |B(v)\rangle \over\sqrt{m_{D_2^*}m_B}}
  &=& k_{A_1}\epsilon^{*\mu\alpha} v_\alpha 
  + (k_{A_2} v^\mu + k_{A_3} v'^\mu)\,
  \epsilon^*_{\alpha\beta} v^\alpha v^\beta  \cr \cr
{\langle D^*_2(v',\epsilon)| \bar c\gamma^\mu b |B(v)\rangle \over\sqrt{m_{D_2^*}m_B}}
  &=& ik_V \varepsilon^{\mu\alpha\beta\gamma} 
  \epsilon^*_{\alpha\sigma} v^\sigma v_\beta v'_\gamma ,  
\end{eqnarray}
where $v~(v')$ is the four-velocity of the $B~(D^{**})$ meson,
$\epsilon^\mu~(\epsilon^{\mu\nu})$ is a polarization vector (tensor) of the final vector
(tensor) charmed meson, and the form factors $f_i$ and $k_i$ are dimensionless 
functions of $w=v\cdot v'$.  The double differential decay rates expressed in terms of
the form factors read as follows \cite{iw1,llsw}:
\begin{eqnarray}\label{ddr1}
{{\rm d}^2\Gamma_{D_1}\over {\rm d}w{\rm d}\!\cos\theta} &=& 
  3\Gamma_0 r_1^3 \sqrt{w^2-1} \bigg\{ \sin^2\theta
  \Big[ (w-r_1) f_{V_1}+(w^2-1) (f_{V_3}+r_1 f_{V_2}) \Big]^2 \cr
&& + (1-2r_1w+r_1^2) \Big[ (1+\cos^2\theta) [f_{V_1}^2 + (w^2-1) f_A^2] 
  - 4\cos\theta \sqrt{w^2-1} f_{V_1} f_A \Big] \bigg\} ,\cr
{{\rm d}^2\Gamma_{D_2^*}\over {\rm d}w{\rm d}\!\cos\theta} &=& 
   \frac32\Gamma_0 r_2^3 (w^2-1)^{3/2} \bigg\{ \frac43\sin^2\theta
  \Big[ (w-r_2) k_{A_1}+(w^2-1) (k_{A_3}+r_2 k_{A_2}) \Big]^2 \cr
&& + (1-2r_2w+r_2^2)\Big[ (1+\cos^2\theta) [k_{A_1}^2 + (w^2-1) k_V^2] 
  - 4\cos\theta \sqrt{w^2-1} k_{A_1} k_V \Big] \bigg\} ,
\end{eqnarray}
where $\Gamma_0 = {G_F^2\,|V_{cb}|^2\,m_B^5 /(192\pi^3)}$, $r_1=m_{D_1}/m_B$,
$r_2=m_{D_2^*}/m_B$, and $\theta$ is the angle between the charged lepton and the 
charmed meson in the rest frame of the virtual $W$ boson. 

The main predictions of HQET for the structure of the $B\to D_1e\nu$ form factors 
look as follows \cite{llsw}:
\begin{eqnarray}\label{sl1}
\sqrt6 f_A &=& - (w+1)\tau 
  - \varepsilon_b \{ (w-1)[(\bar\Lambda'+\bar\Lambda)\tau-(2w+1)\tau_1-\tau_2]+(w+1)\eta_b \} \cr \cr
&& - \varepsilon_c [ 4(w\bar\Lambda'-\bar\Lambda)\tau- 3(w-1) (\tau_1-\tau_2) 
  +(w+1) (\eta_{\rm ke}-2\eta_1-3\eta_3)  ] ,\cr \cr
\sqrt6 f_{V_1} &=&  (1-w^2)\tau 
  - \varepsilon_b (w^2-1) [(\bar\Lambda'+\bar\Lambda)\tau-(2w+1)\tau_1-\tau_2 
+ \eta_b] \cr \cr
&& - \varepsilon_c [ 4(w+1)(w\bar\Lambda'-\bar\Lambda)\tau
-(w^2-1)(3\tau_1-3\tau_2-\eta_{\rm ke}+2\eta_1+3\eta_3) ] , \cr \cr
\sqrt6 f_{V_2} &=&  -3\tau - 3\varepsilon_b [(\bar\Lambda'+\bar\Lambda)\tau-
(2w+1)\tau_1-\tau_2+\eta_b] \cr \cr
&&  - \varepsilon_c [ (4w-1)\tau_1+5\tau_2 +3\eta_{\rm ke} +10\eta_1 
  + 4(w-1)\eta_2-5\eta_3 ] , \cr \cr
\sqrt6 f_{V_3} &=&  (w-2)\tau 
  + \varepsilon_b \{ (2+w)[(\bar\Lambda'+\bar\Lambda)\tau-(2w+1)\tau_1-\tau_2] - (2-w)\eta_b \} \cr \cr
&&  + \varepsilon_c [ 4(w\bar\Lambda'-\bar\Lambda)\tau
+(2+w)\tau_1 + (2+3w)\tau_2  + (w-2)\eta_{\rm ke} \cr \cr
&& - 2(6+w)\eta_1 - 4(w-1)\eta_2 - (3w-2)\eta_3] , 
\end{eqnarray}
where $\varepsilon_Q=1/(2m_Q)$ and $\bar\Lambda(\bar\Lambda')=M(M')-m_Q$ 
is the difference
between the heavy ground state (orbitally excited) meson and heavy quark masses
in the limit $m_Q\to \infty$. The form factor $\tau$ is the leading order Isgur--Wise
function ($\tau$ is $\sqrt3$ times the function $\tau_{3/2}$ of Refs.~\cite{iw1,efg1}).
The subleading Isgur--Wise functions $\tau_1$ and $\tau_2$ originate from the $1/m_Q$
corrections to the $b\to c$ flavour changing current, while $\eta_{ke}$ and $\eta_i$
form factors result from kinetic energy and chromomagnetic corrections to the HQET
Lagrangian.

The analogous formulae for $B\to D_2^*e\nu$ have the form \cite{llsw}
\begin{eqnarray}\label{sl2}
k_V &=& - \tau - \varepsilon_b [(\bar\Lambda'+\bar\Lambda)\tau-(2w+1)\tau_1-\tau_2+\eta_b]
  - \varepsilon_c (\tau_1-\tau_2+\eta_{\rm ke}-2\eta_1+\eta_3) , \cr \cr
k_{A_1} &=& - (1+w)\tau - \varepsilon_b \{ (w-1)[(\bar\Lambda'+\bar\Lambda)\tau-(2w+1)\tau_1-\tau_2]+(1+w)\eta_b \} \cr \cr
&& - \varepsilon_c [ (w-1)(\tau_1-\tau_2)
  + (w+1)(\eta_{\rm ke}-2\eta_1+\eta_3) ] , \cr \cr
k_{A_2} &=& - 2\varepsilon_c (\tau_1+\eta_2) , \cr \cr
k_{A_3} &=& \tau + \varepsilon_b [(\bar\Lambda'+\bar\Lambda)\tau-(2w+1)\tau_1-\tau_2+\eta_b]
  - \varepsilon_c (\tau_1+\tau_2-\eta_{\rm ke}+2\eta_1-2\eta_2-\eta_3) . 
\end{eqnarray}

\subsection{ $B\to D_0^*e\nu$ and $B\to D_1^*e\nu$ decays}

The matrix elements of the vector and axial currents between $B$ mesons and 
$D_0^*$ or $D_1^*$ mesons can be parametrized as follows
\begin{eqnarray}\label{ff2}
\langle D_0^*(v')| \bar c\gamma^\mu b |B(v)\rangle 
  &=& 0, \cr \cr
{\langle D_0^*(v')| \bar c\gamma^\mu\gamma_5 b |B(v)\rangle \over\sqrt{m_{D_0^*}m_B}}
  &=& g_+ (v^\mu+v'^\mu) + g_- (v^\mu-v'^\mu) , \cr \cr
{\langle D_1^*(v',\epsilon)| \bar c\gamma^\mu b |B(v)\rangle \over\sqrt{m_{D_1^*}m_B}}
  &=& g_{V_1} \epsilon^{* \mu} 
  + (g_{V_2} v^\mu + g_{V_3} v'^\mu)\, (\epsilon^*\cdot v) , \cr \cr
{\langle D_1^*(v',\epsilon)| \bar c\gamma^\mu\gamma_5 b |B(v)\rangle \over\sqrt{m_{D_1^*}m_B}}
  &=& i g_A \varepsilon^{\mu\alpha\beta\gamma} 
  \epsilon^*_\alpha v_\beta v'_\gamma ,  
\end{eqnarray}
where the form factors $g_i$ are functions of $w$. In terms of these form factors the double differential
decay rates for $B\to D_0^*\,e\,\bar\nu_e$ and 
$B\to D_1^*\,e\,\bar\nu_e$ decays can be expressed in the following way \cite{llsw}
\begin{eqnarray}\label{ddr2}
{{\rm d}^2\Gamma_{D_0^*}\over {\rm d}w{\rm d}\!\cos\theta} &=& 
  3\Gamma_0 r_0^{*3}\, (w^2-1)^{3/2} \sin^2\theta 
  \Big[ (1+r_0^*)g_+ - (1-r_0^*) g_- \Big]^2 ,\cr \cr
{{\rm d}^2\Gamma_{D_1^*}\over {\rm d}w{\rm d}\!\cos\theta} &=& 
  3\Gamma_0 r_1^{*3} \sqrt{w^2-1} \bigg\{ \sin^2\theta
  \Big[ (w-r_1^*) g_{V_1}+(w^2-1) (g_{V_3}+r_1^* g_{V_2}) \Big]^2 \cr \cr
&& + (1-2r_1^*w+r_1^{*2}) \Big[(1+\cos^2\theta) [g_{V_1}^2 + (w^2-1) g_A^2] 
  - 4\cos\theta \sqrt{w^2-1} g_{V_1} g_A \Big] \bigg\} ,
\end{eqnarray}
where $\Gamma_0 = {G_F^2\,|V_{cb}|^2\,m_B^5 /(192\pi^3)}$, 
$r_0^*=m_{D_0^*}/m_B$ and $r_1^*=m_{D_1^*}/m_B$.
 
The HQET predictions for the form factors of the decay $B\to D_0^*e\nu$ 
are given by \cite{llsw}
\begin{eqnarray}\label{sl3}
g_+ &=& \varepsilon_c \left[ 2(w-1)\zeta_1
  - 3\zeta {w\bar\Lambda^*-\bar\Lambda\over w+1} \right] 
  - \varepsilon_b \left[\frac{\bar\Lambda^*(2w+1)-\bar\Lambda(w+2)}{w+1}\zeta
-2(w-1)\zeta_1\right] , \cr \cr
g_- &=& \zeta + \varepsilon_c \left[ \chi_{\rm ke}+6\chi_1-2(w+1)\chi_2 \right] 
  + \varepsilon_b \chi_b . 
\end{eqnarray}
The analogous formulae for the decay $B\to D_1^*e\nu$ look as follows
\begin{eqnarray}\label{sl4}
g_A &=& \zeta 
  + \varepsilon_c \left[ {w\bar\Lambda^*-\bar\Lambda \over w+1} \zeta
  +\chi_{\rm ke}-2\chi_1 \right] -\varepsilon_b\left[\frac{\bar\Lambda^*(2w+1)-\bar\Lambda(w+2)}{w+1}\zeta
-2(w-1)\zeta_1-\chi_b\right] ,\cr \cr
g_{V_1} &=&  (w-1)\zeta + \varepsilon_c 
  \left[(w\bar\Lambda^*-\bar\Lambda)\zeta + (w-1)(\chi_{\rm ke}-2\chi_1) \right] 
  \cr \cr
&& - \varepsilon_b \left\{ [\bar\Lambda^*(2w+1)-\bar\Lambda(w+2)]
\zeta-2(w^2-1)\zeta_1 - (w-1)\chi_b \right\} ,\cr \cr
g_{V_2} &=& 2\varepsilon_c (\zeta_1-\chi_2) , \cr \cr
g_{V_3} &=& -\zeta 
  - \varepsilon_c \left[ {w\bar\Lambda^*-\bar\Lambda \over w+1}\zeta 
  + 2\zeta_1 + \chi_{\rm ke} - 2\chi_1 +2\chi_2 \right]\cr \cr 
 && + \varepsilon_b \left[\frac{\bar\Lambda^*(2w+1)-\bar\Lambda(w+2)}{w+1}\zeta-2(w-1)\zeta_1-\chi_b
\right] . 
\end{eqnarray}
The form factor $\zeta$ is the leading order Isgur--Wise
function ($\zeta$ is twice the function $\tau_{1/2}$ of Refs.~\cite{iw1,efg1}).
The subleading Isgur--Wise function $\zeta_1$  originates from the $1/m_Q$
corrections to the $b\to c$ flavour changing current, while $\chi_{ke}$ and $\chi_i$
form factors result from kinetic energy and chromomagnetic corrections to the HQET
Lagrangian.

In the following sections we apply the relativistic quark model to the calculation
of leading and subleading Isgur--Wise functions.

\section{Relativistic quark model}

In the quasipotential approach, a meson is described by the wave
function of the bound quark--antiquark state, which satisfies the
quasipotential equation \cite{3} of the Schr\"odinger type~\cite{4}:
\begin{equation}
\label{quas}
{\left(\frac{b^2(M)}{2\mu_{R}}-\frac{{\bf
p}^2}{2\mu_{R}}\right)\Psi_{M}({\bf p})} =\int\frac{d^3 q}{(2\pi)^3}
 V({\bf p,q};M)\Psi_{M}({\bf q}),
\end{equation}
where the relativistic reduced mass is
\begin{equation}
\mu_{R}=\frac{M^4-(m^2_q-m^2_Q)^2}{4M^3}.
\end{equation}
Here $m_{q,Q}$ are the masses of light
and heavy quarks, and ${\bf p}$ is their relative momentum.  
In the centre-of-mass system the relative momentum squared on mass shell 
reads
\begin{equation}
{b^2(M) }
=\frac{[M^2-(m_q+m_Q)^2][M^2-(m_q-m_Q)^2]}{4M^2}.
\end{equation}

The kernel 
$V({\bf p,q};M)$ in Eq.~(\ref{quas}) is the quasipotential operator of
the quark--antiquark interaction. It is constructed with the help of the
off-mass-shell scattering amplitude, projected onto the positive
energy states. An important role in this construction is played 
by the Lorentz-structure of the confining quark--antiquark interaction
in the meson.  In 
constructing the quasipotential of the quark-antiquark interaction 
we have assumed that the effective
interaction is the sum of the usual one-gluon exchange term and the mixture
of vector and scalar linear confining potentials.
The quasipotential is then defined by
\cite{mass}  
\begin{eqnarray}
\label{qpot}
V({\bf p,q};M)&=&\bar{u}_q(p)\bar{u}_Q(-p){\cal V}({\bf p},{\bf q};M)u_q(q)u_Q(-q)\cr \cr
&=&\bar{u}_q(p)
\bar{u}_Q(-p)\Bigg\{\frac{4}{3}\alpha_sD_{ \mu\nu}({\bf
k})\gamma_q^{\mu}\gamma_Q^{\nu}\cr\cr
& & +V^V_{\rm conf}({\bf k})\Gamma_q^{\mu}
\Gamma_{Q;\mu}+V^S_{\rm conf}({\bf
k})\Bigg\}u_q(q)u_Q(-q),
\end{eqnarray}
where $\alpha_s$ is the QCD coupling constant, $D_{\mu\nu}$ is the
gluon propagator in the Coulomb gauge
and ${\bf k=p-q}$; $\gamma_{\mu}$ and $u(p)$ are 
the Dirac matrices and spinors
\begin{equation}
\label{spinor}
u^\lambda({p})=\sqrt{\frac{\epsilon(p)+m}{2\epsilon(p)}}
{1\choose \frac{\bbox{\sigma p}}{\epsilon(p)+m}}\chi^\lambda
\end{equation}
with $\epsilon(p)=\sqrt{{\bf p}^2+m^2}$.
The effective long-range vector vertex is
given by
\begin{equation}
\Gamma_{\mu}({\bf k})=\gamma_{\mu}+
\frac{i\kappa}{2m}\sigma_{\mu\nu}k^{\nu},
\end{equation}
where $\kappa$ is the Pauli interaction constant characterizing the
anomalous chromomagnetic moment of quarks. Vector and
scalar confining potentials in the non-relativistic limit reduce to
\begin{equation}
V^V_{\rm conf}(r)=(1-\varepsilon)(Ar+B),\qquad
V^S_{\rm conf}(r) =\varepsilon (Ar+B),
\end{equation}
reproducing 
\begin{equation}
V_{\rm conf}(r)=V^S_{\rm conf}(r)+
V^V_{\rm conf}(r)=Ar+B,
\end{equation}
where $\varepsilon$ is the mixing coefficient. 

The quasipotential for the heavy quarkonia,
expanded in $v^2/c^2$, can be found in Refs.~\cite{mass,pot} and for
heavy--light mesons in \cite{egf}.
All the parameters of
our model, such as quark masses, parameters of the linear confining potential,
mixing coefficient $\varepsilon$ and anomalous
chromomagnetic quark moment $\kappa$, were fixed from the analysis of
heavy quarkonia masses \cite{mass} and radiative decays \cite{gf}. 
The quark masses
$m_b=4.88$ GeV, $m_c=1.55$ GeV, $m_s=0.50$ GeV, $m_{u,d}=0.33$ GeV and
the parameters of the linear potential $A=0.18$ GeV$^2$ and $B=-0.30$ GeV
have usual quark model values.  The value of the vector-scalar mixing
coefficient $\varepsilon=-1$
has been determined by considering the heavy quark expansion
\cite{fg} and meson radiative decays \cite{gf}.
Finally, the universal Pauli interaction constant $\kappa=-1$ has been
fixed from the analysis of the fine splitting of heavy quarkonia ${
}^3P_J$- states \cite{mass}. Note that the 
long-range  magnetic contribution to the potential in our model
is proportional to $(1+\kappa)$ and thus vanishes for the 
chosen value of $\kappa=-1$.

In order to calculate the exclusive semileptonic decay rate of the $B$ meson, 
it is necessary to determine the corresponding matrix element of the 
weak current between meson states.
In the quasipotential approach, the matrix element of the weak current $J^W=\bar c\gamma_\mu(1-\gamma^5)b$ 
between a $B$ meson and an orbitally excited $D^{**}$ meson takes
 the form \cite{f}
\begin{equation}\label{mxet} 
\langle D^{**} \vert J^W_\mu (0) \vert B\rangle
=\int \frac{d^3p\, d^3q}{(2\pi )^6} \bar \Psi_{D^{**}}({\bf
p})\Gamma _\mu ({\bf p},{\bf q})\Psi_B({\bf q}),\end{equation}
where $\Gamma _\mu ({\bf p},{\bf
q})$ is the two-particle vertex function and  $\Psi_{B,D^{**}}$ are the
meson wave functions projected onto the positive energy states of
quarks and boosted to the moving reference frame.
 The contributions to $\Gamma$ come from Figs.~1 and 2.\footnote{  
The contribution $\Gamma^{(2)}$ is the consequence
of the projection onto the positive-energy states. Note that the form of the
relativistic corrections resulting from the vertex function
$\Gamma^{(2)}$ is explicitly dependent on the Lorentz structure of the
$q\bar q$-interaction.} In the heavy quark limit
$m_{b,c}\to \infty$ only $\Gamma^{(1)}$ contributes, while $\Gamma^{(2)}$ 
contributes at $1/m_{Q}$ order. 
They look like
\begin{equation} \label{gamma1}
\Gamma_\mu^{(1)}({\bf
p},{\bf q})=\bar u_{c}(p_c)\gamma_\mu(1-\gamma^5)u_b(q_b)
(2\pi)^3\delta({\bf p}_q-{\bf
q}_q),\end{equation}
and
\begin{eqnarray}\label{gamma2} 
\Gamma_\mu^{(2)}({\bf
p},{\bf q})&=&\bar u_{c}(p_c)\bar u_q(p_q) \Bigl\{\gamma_{Q\mu}(1-\gamma_Q^5)
\frac{\Lambda_b^{(-)}(
k)}{\epsilon_b(k)+\epsilon_b(p_c)}\gamma_Q^0
{\cal V}({\bf p}_q-{\bf
q}_q)\nonumber \\ 
& &+{\cal V}({\bf p}_q-{\bf
q}_q)\frac{\Lambda_{c}^{(-)}(k')}{ \epsilon_{c}(k')+
\epsilon_{c}(q_b)}\gamma_Q^0 \gamma_{Q\mu}(1-\gamma_Q^5)\Bigr\}u_b(q_b)
u_q(q_q),\end{eqnarray}
where the superscripts ``(1)" and ``(2)" correspond to Figs.~1 and
2, $Q= c$ or $b$, ${\bf k}={\bf p}_c-{\bf\Delta};\
{\bf k}'={\bf q}_b+{\bf\Delta};\ {\bf\Delta}={\bf
p}_{D^{**}}-{\bf p}_B; \ \epsilon (p)=(m^2+{\bf p}^2)^{1/2}$;
$$\Lambda^{(-)}(p)=\frac{\epsilon(p)-\bigl( m\gamma
^0+\gamma^0({\bbox{ \gamma p}})\bigr)}{ 2\epsilon (p)}.$$
Here \cite{f} 
\begin{eqnarray*} 
p_{c,q}&=&\epsilon_{c,q}(p)\frac{p_{D^{**}}}{M_{D^{**}}}
\pm\sum_{i=1}^3 n^{(i)}(p_{D^{**}})p^i,\\
q_{b,q}&=&\epsilon_{b,q}(q)\frac{p_B}{M_B} \pm \sum_{i=1}^3 n^{(i)}
(p_B)q^i,\end{eqnarray*}
and $n^{(i)}$ are three four-vectors given by
$$ n^{(i)\mu}(p)=\left\{ \frac{p^i}{M},\ \delta_{ij}+
\frac{p^ip^j}{M(E+M)}\right\}, \quad E=\sqrt{{\bf p}^2+M^2}.$$

The wave function of a $P$-wave $D^{**}$ meson at rest is given by
\begin{equation}\label{psi}
\Psi_{D^{**}}({\bf p})\equiv
\Psi^{JM}_{D(j)}({\bf p})={\cal Y}^{JM}_j\psi_{D(j)}({\bf p}),
\end{equation}
where $J$ and $M$ are the total meson angular momentum and its projection,
while $j$ is the light quark angular momentum;   
$\psi_{D(j)}({\bf p})$ is the radial part of the wave function,
which has been determined by the numerical solution of Eq.~(\ref{quas})
in \cite{egf}.
The spin-angular momentum part ${\cal Y}^{JM}_j$ has the following form
\begin{eqnarray}\label{angl}
{\cal Y}^{JM}_j&=&\sum_{\sigma_Q\sigma_q}\left\langle j\, M-\sigma_Q,\  
\frac12\, \sigma_Q |J\, M\right\rangle\left\langle 1\, M-\sigma_Q-\sigma_q,\ 
\frac12\, \sigma_q |j\, M-\sigma_Q\right\rangle \cr \cr
 & &\times Y_{1}^{M-\sigma_Q-\sigma_q}
\chi_Q(\sigma_Q)\chi_q(\sigma_q).
\end{eqnarray}
Here $\langle j_1\, m_1,\  j_2\, m_2|J\, M\rangle$ are Clebsch-Gordan 
coefficients, $Y_l^m$ are spherical harmonics, and $\chi(\sigma)$ (where 
$\sigma=\pm 1/2$) are spin wave functions :
$$ \chi\left(1/2\right)={1\choose 0}, \qquad 
\chi\left(-1/2\right)={0\choose 1}. $$

It is important to note that the wave functions entering the weak current
matrix element (\ref{mxet}) are not in the rest frame in general. For example, 
in the $B$ meson rest frame, the $D^{**}$ meson is moving with the recoil
momentum ${\bf \Delta}$. The wave function
of the moving $D^{**}$ meson $\Psi_{D^{**}\,{\bf\Delta}}$ is connected 
with the $D^{**}$ wave function in the rest frame $\Psi_{D^{**}\,{\bf 0}}\equiv
\Psi_{D(j)}$ by the transformation \cite{f}
\begin{equation}
\label{wig}
\Psi_{D^{**}\,{\bf\Delta}}({\bf
p})=D_c^{1/2}(R_{L_{\bf\Delta}}^W)D_q^{1/2}(R_{L_{
\bf\Delta}}^W)\Psi_{D^{**}\,{\bf 0}}({\bf p}),
\end{equation}
where $R^W$ is the Wigner rotation, $L_{\bf\Delta}$ is the Lorentz boost
from the meson rest frame to a moving one, and   
the rotation matrix $D^{1/2}(R)$ in spinor representation is given by
\begin{equation}\label{d12}
{1 \ \ \,0\choose 0 \ \ \,1}D^{1/2}_{c,q}(R^W_{L_{\bf\Delta}})=
S^{-1}({\bf p}_{c,q})S({\bf\Delta})S({\bf p}),
\end{equation}
where
$$
S({\bf p})=\sqrt{\frac{\epsilon(p)+m}{2m}}\left(1+\frac{\bbox{ \alpha p}}
{\epsilon(p)+m}\right)
$$
is the usual Lorentz transformation matrix of the four-spinor.
For electroweak $B$ meson
decays to $S$-wave final mesons such a transformation contributes at first
order of the $1/m_Q$ expansion, while for the decays to excited final mesons
it gives a contribution already to the leading term due to the orthogonality
of the initial and final meson wave functions.

\section{Leading and subleading Isgur--Wise functions}

Now we can perform the heavy quark expansion for the matrix elements
of $B$ decays to orbitally excited $D$ mesons in the framework of our model and
determine leading and subleading Isgur--Wise functions. We substitute
the vertex functions $\Gamma^{(1)}$  and $\Gamma^{(2)}$ 
given by Eqs.~(\ref{gamma1}) and (\ref{gamma2})
in the decay matrix element (\ref{mxet}) and take into account the wave function
properties (\ref{psi})--(\ref{wig}).~\footnote{Note
that the quark model definition of $\Psi_{D^*_1(1/2)}$ in (\ref{psi}), (\ref{angl})
differs from the HQET one \cite{iw1,llsw}  by an overall minus sign.}
The resulting structure of this matrix element is
rather complicated, because it is necessary to integrate both over  $d^3 p$
and $d^3 q$. The $\delta$ function in expression (\ref{gamma1}) permits us to perform
one of these integrations and thus this contribution can be easily calculated. The
calculation  
of the vertex function $\Gamma^{(2)}$ contribution is more difficult. Here, instead
of a $\delta$ function, we have a complicated structure, containing the 
$Q\bar q$ interaction potential in the meson. 
However, we can expand this contribution in inverse
powers of heavy ($b,c$) quark masses and then use the quasipotential equation in
order to perform one of the integrations in the current matrix element. We carry out 
the heavy quark expansion up to first order in $1/m_Q$. It is easy to see that the vertex
function $\Gamma^{(2)}$ contributes already at  the subleading order of the $1/m_Q$
expansion. Then we compare the arising  decay matrix elements with
the form factor decompositions (\ref{ff1}) and (\ref{ff2}) and determine the corresponding
form factors. We find that, for the chosen values of our model parameters (the mixing
coefficient of vector and scalar confining potential $\varepsilon=-1$ and the Pauli
constant $\kappa=-1$), the resulting structure  at leading
and subleading order in $1/m_Q$ coincides with the model-independent predictions
of HQET given by Eqs.~(\ref{sl1}), (\ref{sl2}), (\ref{sl3}), and (\ref{sl4}). 
We get the following
expressions for leading and subleading Isgur--Wise functions:

i) $B\to D_1e\nu$ and $B\to D^*_2e\nu$ decays
\begin{eqnarray}
\label{tau}
\tau(w)&=&\sqrt{\frac23}\frac{1}{(w+1)^{3/2}}\int\frac{d^3 p}
{(2\pi)^3}\bar\psi_{D(3/2)}\left({\bf p}+\frac{2\epsilon_q}{M_{D(3/2)}(w+1)}
{\bf \Delta}\right)\cr
&&\times\left[-2\epsilon_q
\overleftarrow{\frac{\partial}{\partial p}}+\frac{p}{\epsilon_q+m_q}\right]
\psi_B({\bf p}),\\
\label{tau1}
\tau_1(w)&=&\frac{\bar\Lambda'+\bar\Lambda}{w+1}\tau(w),\\
\label{tau2}
\tau_2(w)&=&-\frac{w}{w+1}(\bar\Lambda'+\bar\Lambda)\tau(w).
\end{eqnarray}

ii) $B\to D^*_0e\nu$ and $B\to D^*_1e\nu$ decays
\begin{eqnarray}
\label{zeta}
\zeta(w)&=&\frac{\sqrt{2}}{3}\frac{1}{(w+1)^{1/2}}\int\frac{d^3 p}
{(2\pi)^3}\bar\psi_{D(1/2)}\left({\bf p}+\frac{2\epsilon_q}{M_{D(1/2)}(w+1)}
{\bf \Delta}\right)\cr
&&\times\left[-2\epsilon_q
\overleftarrow{\frac{\partial}{\partial p}}-\frac{2p}{\epsilon_q+m_q}\right]
\psi_B({\bf p}),\\
\label{zeta1}
\zeta_1(w)&=&\frac{\bar\Lambda^*+\bar\Lambda}{w+1}\zeta(w).
\end{eqnarray}
The contributions of all other subleading form factors, $\eta_i(w)$ and 
$\chi_i(w)$, to decay matrix elements are suppressed by an additional
power of the ratio $(w-1)/(w+1)$, which is equal to zero at $w=1$ and less than
$1/6$ at $w_{\rm max}=(1+r^2)/(2r)$ ($r=r_1, r_2, r^*_0,$ or $r^*_1$ respectively).
Since the main contribution
to the decay rate comes from the values of form factors close to $w=1$, these form
factors turn out to be unimportant. This result is in agreement with the HQET-motivated 
considerations \cite{llsw} that the functions parametrizing the time-ordered products 
of the chromomagnetic term in the HQET Lagrangian with the leading
order currents should be small.     

The arrow over $\partial/\partial p$ in (\ref{tau}) and (\ref{zeta}) indicates that the
derivative acts on the wave function of the $D^{**}$ meson. All the wave functions and
meson masses have been obtained in \cite{egf} by the numerical solution of the 
quasipotential equation. We use the following values for HQET parameters 
$\bar\Lambda=0.51$~GeV, $\bar\Lambda'=0.80$~GeV, and 
$\bar\Lambda^*= 0.89$~GeV \cite{egf}.

The last  terms in 
the square brackets of the expressions for the leading order Isgur--Wise functions 
$\tau(w)$ (\ref{tau}) and $\zeta(w)$ (\ref{zeta}) result from the wave 
function transformation (\ref{wig}) 
associated with the relativistic  rotation of the
light quark spin (Wigner rotation) in
passing to the moving reference frame. These terms are numerically important
and lead to the suppression of the $\zeta$ form factor compared to 
$\tau$. Note that if we  
had applied a simplified non-relativistic quark model \cite{iw1,vo}
these important contributions would be missing. Neglecting further the
small difference between the wave functions $\psi_{D(1/2)}$ and 
$\psi_{D(3/2)}$, the following relation between $\tau$ and 
$\zeta$ would have been obtained \cite{llsw}
\begin{equation}\label{taunr}
\zeta(w)=\frac{w+1}{\sqrt{3}}\tau(w). 
\end{equation}
However, we see that this relation is violated if the relativistic
transformation properties of the wave function are taken into account. 
At the point $w=1$, where the initial $B$ meson and final $D^{**}$ are
at rest, we find instead the relation
\begin{equation}\label{diftau}
\frac{\tau(1)}{\sqrt{3}}-\frac{\zeta(1)}{2}\cong \frac12\int\frac{d^3p}{(2\pi)^3}
\bar\psi_{D^{**}}({\bf p})\frac{p}{\epsilon_q+m_q}\psi_B({\bf p}),
\end{equation}
obtained by assuming $\psi_{D(3/2)}\cong\psi_{D(1/2)}\cong\psi_{D^{**}}$.
The relation (\ref{diftau}) coincides with the one found in 
Ref.~\cite{mlopr}, where the Wigner rotation was also taken into account. 

\section{Numerical results and predictions}
In Table~\ref{tauv} we present our numerical results for the leading order
Isgur--Wise functions $\tau(1)$ and $\zeta(1)$ at zero recoil of the final $D^{**}$ meson, 
as well as their slopes $\left.\rho_{3/2}^2=-\frac{1}{\tau}\frac{\partial}{\partial w}
\tau\right|_{w=1}$ and $\left.\rho_{1/2}^2=-\frac{1}{\zeta}\frac{\partial}{\partial w}
\zeta\right|_{w=1}$,
in comparison with other model predictions 
\cite{llsw,mlopr,ddgnp,w,cdp,gi,cccn}.  We see that most of 
the above approaches predict close values for the function $\tau(1)$ 
and its slope $\rho_{3/2}^2$, while the results for $\zeta(1)$ 
significantly differ from one another. This difference is a consequence
of a specific treatment of the relativistic quark dynamics. Non-relativistic
approaches predict $\zeta(1)\simeq(2/\sqrt{3})\tau(1)$ (see (\ref{taunr})),
while the relativistic treatment leads to $(2/\sqrt{3})\tau(1)>\zeta(1)$
(see Eq.~(\ref{diftau})). The more relativistic the light quark in the heavy--light
meson is, the more suppressed  $\zeta$ is with respect to $\tau$.
We plot our results for leading ($\tau(w)$, $\zeta(w)$) and subleading ($\tau_1(w)$,
$\tau_2(w)$, $\zeta_1(w)$) Isgur--Wise functions for $B\to D^{**}e\nu$ in Figs.~3,~4 
and for  $B_s\to D^{**}_s e\nu$ in Figs.~5,~6.   

We can now calculate the decay branching ratios by integrating double differential  decay 
rates in Eqs.~(\ref{ddr1}) and (\ref{ddr2}). Our results for decay rates both in
the infinitely heavy 
quark limit and taking account of the first order $1/m_Q$ corrections as well as
their ratio
$$R=\frac{{\rm Br}(B\to D^{**}e\nu)_{{\rm with}\, 1/m_Q}}{{\rm Br}(B\to
D^{**}e\nu)_{m_Q\to\infty}}$$
are presented in Tables~\ref{bd} and \ref{bds}. We see that the inclusion of $1/m_Q$
corrections considerably influences the results and for some decays their contribution
is as important as the leading order contribution. This is the consequence of the 
vanishing of the leading order contribution to the decay matrix elements due to the
heavy quark spin-flavour symmetry at zero recoil of the final $D^{**}$ meson, while 
nothing prevents $1/m_Q$ corrections to contribute to the decay matrix element at 
this kinematical point. In fact, from Eqs.~(\ref{ff1}) and (\ref{ff2}), we see that  decay
matrix elements at zero recoil are determined by form factors $f_{V_1}(1)$, 
$g_+(1)$ and $g_{V_1}(1)$, which receive non-vanishing contributions from
first order heavy quark mass corrections. From Eqs.~(\ref{sl1}), (\ref{sl3}), and
(\ref{sl4}) we find
\begin{eqnarray}
\label{fv1}
\sqrt{6} f_{V_1}(1)&=& -8\varepsilon_c(\bar\Lambda'-\bar\Lambda)\tau(1)\\
\label{g+}
g_+(1)&=&-\frac32(\varepsilon_c+\varepsilon_b)(\bar\Lambda^*-\bar\Lambda)
\zeta(1)\\
\label{gv1}
g_{V_1}(1)&=&(\varepsilon_c-3\varepsilon_b)(\bar\Lambda^*-\bar\Lambda)
\zeta(1).
\end{eqnarray}
Since the kinematically allowed range for these decays is not broad ($1\le w\le 
w_{\rm max}\approx 1.32$), the contribution to the decay rate of the 
rather small $1/m_Q$ corrections is substantially increased. This is confirmed by
numerical calculations. From Tables~\ref{bd} and \ref{bds} we see that the decay
rate $B\to D^*_2e\nu$, for which all contributions vanish at zero recoil, is only slightly
increased by subleading $1/m_Q$ corrections. On the other hand, $B\to D_1e\nu$
and $B\to D^*_0e\nu$ decay rates receive large $1/m_Q$ contributions. The situation
is different for the $B\to D^*_1e\nu$ decay. Here the $1/m_Q$ contribution at zero recoil
is not equal to zero, but it is suppressed  by a very small factor $(\varepsilon_c-
3\varepsilon_b)$ (see Eq.~(\ref{gv1})), which is only $\approx 0.03$~GeV$^{-1}$ for our
model parameters. As a result the $B\to D^*_1e\nu$ decay rate receives $1/m_Q$ 
contributions comparable to those for the $B\to D^*_2e\nu$ rate. The above discussion shows 
that the sharp increase of $B\to D_1e\nu$ and $B\to D^*_0$ decay rates
by first order $1/m_Q$ corrections does not
signal the breakdown of the heavy quark expansion, but is rather a result of the
interplay of kinematical and dynamical effects. Thus we have  good reasons
to expect that higher order $1/m_Q$ corrections will influence these decay rates
at the level of 10 -- 20 \%.  

In Table~\ref{bd} we present the experimental data from CLEO \cite{cleo} and
ALEPH \cite{aleph}, which are available only for the $B\to D_1 e\nu$ decay. For
$B\to D^*_2e\nu$, these experimental groups present only upper limits, which
require the use of some additional assumptions about the hadronic branching 
ratios of the
$D^*_2$ meson. Our result for the branching ratio of the $B\to D_1e\nu$ decay with
the inclusion of $1/m_Q$ corrections is in good agreement with both measurements.
On the other hand, our branching ratio for the $B\to D^*_2e\nu$ decay is 
only within the CLEO upper
limit and disagrees with the ALEPH one. However, there are some reasons to expect 
that the ALEPH bound is too strong \cite{llsw}. 

 In Table~\ref{br} we present our
predictions for the ratios of decay rates $B\to D^*_2e\nu$, $B\to D^*_0e\nu$,
$B\to D^*_1e\nu$, and of the sum of all $B\to D^{**}e\nu$ decay rates to the rate 
$B\to D_1e\nu$  both in the limit  $m_Q\to \infty$, and
taking into account the $1/m_Q$ corrections. It is reasonable to consider such 
ratios in order to normalize to a measured rate.  In Ref.~\cite{mlopr} it is 
argued that a ratio ${\rm Br}(B\to D^*_2e\nu)/{\rm Br}(B\to D_1e\nu)=1.55\pm 0.15$ is a
mere consequence of the heavy quark symmetry. In the heavy quark limit
we confirm this result. However, the inclusion of $1/m_Q$ corrections strongly
influences this prediction and considerably reduce this ratio to a value close to 1.
Such a reduction seems to be favoured by available experimental data. In the last row of 
Table~\ref{br} we give the sum of all $B\to D^{**}e\nu$ branching ratios.
We see that our model predicts that 1.45\% of $B$ meson decays go to the 
first orbitally excited $D$ meson states. This result means that approximately     
2.5\% of $B$ decays should go to higher excitations.

In Fig.~7 we plot the electron spectra $(1/\Gamma_0)({\rm d}\Gamma/{\rm d}y)$ for
$B\to D^{**}e\nu$ decays. Here $y=2E_e/m_B$ is the rescaled lepton energy. These
differential decay rates can be easily obtained from double differential decay rates
(\ref{ddr1}), (\ref{ddr2}), using the relation
$y=1-rw-r\sqrt{w^2-1}\cos\theta$
and then integrating in $w$ over $[(1-y)^2+r^2]/[2r(1-y)]<w<(1+r^2)/(2r)$. We present
our results both in the heavy quark limit $m_Q\to\infty$ (dashed curves) and with the
inclusion of first order $1/m_Q$ corrections (solid curves).

\section{Bjorken sum rule}
Finally we test the fulfilment of the Bjorken 
sum rule \cite{b} in our model.
This sum rule states
\begin{equation}
\label{bsr}
\rho^2=\frac14+\sum_m\frac{|\zeta^{(m)}(1)|^2}{4}
 +2\sum_m \frac{|\tau^{(m)}(1)|^2}{3}
+\cdots ,
\end{equation}
where $\rho^2$ is the slope of the $B\to D^{(*)}e\nu$ Isgur--Wise function,
$\zeta^{(m)}$ and $\tau^{(m)}$  are the form factors describing the
orbitally excited states discussed here and their
radial excitations, and  
ellipses denote contributions from non-resonant channels. 
We see that the contribution of the lowest lying $P$-wave
states  implies the bound 
\begin{equation}
\rho^2>\frac14 +\frac{|\zeta(1)|^2}{4}+2\frac{|\tau(1)|^2}{3}=0.81,
\end{equation}  
which is in agreement with the slope $\rho^2=1.02$ in our model \cite{fg}
and with experimental values \cite{pdg}.

\section{Conclusions}
In this paper we have applied the relativistic quark model to the consideration
of semileptonic $B$ decays to orbitally excited charmed mesons, in the leading
and subleading order of the heavy quark expansion.  We have found an interesting 
interplay of the relativistic and finite heavy quark mass contributions. In particular,
it has been found that  the Lorentz transformation properties of meson wave functions 
play an important role in the theoretical description of these decays. 
Thus, the Wigner rotation of the light quark spin gives a significant contribution 
already at the leading order of the heavy quark expansion. This contribution
considerably reduces the leading order Isgur--Wise function $\zeta$ with respect to
$\tau$. As a result, in this limit, the decay rates $B\to D^*_0e\nu$ and $B\to D^*_1e\nu$     
are approximately an order of magnitude smaller than the decay rates $B\to D_1e\nu$ and 
$B\to D^*_2e\nu$. On the other hand, inclusion of the first order $1/m_Q$ corrections
also substantially influences the decay rates. This large effect of subleading heavy
quark corrections is a consequence of the vanishing of the leading order contributions
to the decay matrix elements due to heavy quark spin-flavour symmetry at the point
of zero recoil of the final charmed meson. However, the subleading order contributions
to $B\to D_1e\nu$, $B\to D^*_0e\nu$ and $B\to D^*_1e\nu$ do not vanish at this
kinematical point. Since the kinematical range for these decays is rather small, the
role of these corrections is considerably increased. Their account results in
an approximately twofold enhancement of the $B\to D_1e\nu$ and $B\to D^*_0e\nu$
decay rates, while the  $B\to D^*_2e\nu$ and $B\to D^*_1e\nu$ rates are 
increased only slightly. The small influence of $1/m_Q$ corrections on the
$B\to D^*_1e\nu$ decay rate is the consequence of the additional interplay of
$1/m_c$ and $1/m_b$ corrections at the zero recoil point (see Eq.~(\ref{gv1})). We thus see
that these subleading heavy quark corrections turn out to be very important
and considerably change the infinitely heavy quark limit results. 
For example, the ratio
of branching ratios ${\rm Br}(B\to D^*_2e\nu)/{\rm Br}(B\to D_1e\nu)$ changes from
the value of about 1.6 in the heavy quark limit, $m_Q\to\infty$,
to the value of about 1 after subleading corrections are included.

In conclusion, we have presented here the first self-consistent dynamical calculation
of subleading heavy quark corrections in the framework of the relativistic quark
model, which are found to be in agreement with the HQET predictions.

\acknowledgements
We thank D.V. Antonov, M. Beneke,  J.G. K\"orner and V.I. Savrin 
for useful discussions.
R.N.F. gratefully acknowledges the warm hospitality
of the colleagues in the particle theory group of the Humboldt-University
extended to him during his stay there.
R.N.F and V.O.G. were supported in part by the {\it Deutsche
Forschungsgemeinschaft} under contract Eb 139/1-3.

\begin{table}
\caption{The comparison of our results for
the values of the leading Isgur--Wise functions $\tau$ and $\zeta$ at zero 
recoil of the final $D^{**}$ meson  and their slopes $\rho_j^2$ with other
predictions.} 
\label{tauv}
\begin{tabular}{cccccccc}
   & Ours & \cite{llsw} & \cite{ddgnp} & \cite{w} & \cite{cdp}& 
\cite{mlopr},\cite{gi} & \cite{mlopr},\cite{cccn}\\
\hline
$\tau(1)$ & 0.85 & 0.71 & 0.97 & 1.14 &     & 1.02 & 0.90\\
$\rho_{3/2}^2$  & 1.53 & 1.5  & 2.3  & 1.9  &     & 1.5  & 1.45\\
$\zeta(1)$ & 0.59 & 0.82 & 0.18 & 0.82 &$0.70\pm0.16$ & 0.44 & 0.12\\
$\rho_{1/2}^2$  & 1.04 & 1.0  & 1.1  & 1.4  &$2.5\pm1.0$ & 0.83 & 0.73\\ 
\end{tabular}
\end{table}
    
\begin{table}
\caption{Decay rates $\Gamma$ (in units of $|V_{cb}/0.04|^2\times 10^{-15}$ GeV) 
and branching ratios BR (in \%) for $B\to D^{**}e\nu$ decays in the infinitely heavy quark
mass limit and taking account of first order $1/m_Q$ corrections. $R$ is a ratio
of branching ratios taking account of $1/m_Q$ corrections to branching ratios in
the infinitely heavy quark mass limit.  }
\label{bd}
\begin{tabular}{cccccccc}
   &\multicolumn{2}{c}{$m_Q\to\infty$}&\multicolumn{2}{c}{With $1/m_Q$}&
&\multicolumn{2}{c}{Experiment}\\
Decay& $\Gamma$ & Br& $\Gamma$ & Br &$R$& Br (CLEO) \cite{cleo}  
& Br (ALEPH) \cite{aleph} \\
\hline
$B\to D_1e\nu$&1.4&0.32  & 2.7 & 0.63& 1.97
& $0.56\pm 0.13\pm0.08\pm0.04$& $0.74\pm0.16$\\
$B\to D^*_2e\nu$&2.1&0.51 & 2.5 & 0.59& 1.16& $<0.8$ & $<0.2$\\
$B\to D^*_1e\nu$&0.31&0.073 & 0.39 & 0.09&1.23 & &\\
$B\to D^*_0e\nu$&0.25&0.061 & 0.59 & 0.14&2.3 & &\\
\end{tabular}
\end{table}

\begin{table}
\caption{Decay rates $\Gamma$ (in units of $|V_{cb}/0.04|^2\times 10^{-15}$ GeV) 
and branching ratios BR (in \%) for $B\to D_s^{**}e\nu$ decays in the infinitely heavy quark
mass limit and taking account of first order $1/m_Q$ corrections. $R$ is a ratio
of branching ratios taking account of $1/m_Q$ corrections to branching ratios in
the infinitely heavy quark mass limit.  }
\label{bds}
\begin{tabular}{cccccc}
   &\multicolumn{2}{c}{$m_Q\to\infty$}&\multicolumn{2}{c}{With $1/m_Q$}\\
Decay& $\Gamma$ & Br& $\Gamma$ & Br &$R$ \\
\hline
$B\to D_{s1}e\nu$&1.5&0.36  & 4.5 & 1.06& 2.9\\
$B\to D^*_{s2}e\nu$&2.4&0.56 & 3.2 & 0.75& 1.3\\
$B\to D^*_{s1}e\nu$&0.53&0.13 & 0.77 & 0.18&1.4 \\
$B\to D^*_{s0}e\nu$&0.44&0.10 & 1.6 & 0.37&3.6 \\
\end{tabular}
\end{table} 

\begin{table}
\caption{Predictions for  ratios of decay rates $B\to D^*_2e\nu$, $B\to D^*_0e\nu$,
and $B\to D^*_1e\nu$ to the rate $B\to D_1e\nu$ in the $m_Q\to \infty$ limit and
taking account of $1/m_Q$ corrections. In the last line we show our predictions
for the sum of branching ratios (in \%) of $B$ decays to orbitally excited $D^{**}$ 
mesons.}
\label{br}
\begin{tabular}{ccc}
 & $m_Q\to\infty$ & With $1/m_Q$\\
\hline
${{\rm Br}(B\to D^*_2e\nu)}/{{\rm Br}(B\to D_1e\nu)}$ & 1.59 &0.94\\
${{\rm Br}(B\to D^*_0e\nu)}/{{\rm Br}(B\to D_1e\nu)}$ & 0.19 &0.22\\
${{\rm Br}(B\to D^*_1e\nu)}/{{\rm Br}(B\to D_1e\nu)}$ & 0.23 &0.14\\
${\sum{\rm Br}(B\to D^{**}e\nu)}/{{\rm Br}(B\to D_1e\nu)}$ & 3.0& 2.3\\
\hline
$\sum {\rm Br}(B\to D^{**}e\nu)$ & 0.96& 1.45\\
\end{tabular}
\end{table}

\begin{figure}
\unitlength=0.9mm
\large
\begin{picture}(150,150)
\put(10,100){\line(1,0){50}}
\put(10,120){\line(1,0){50}}
\put(35,120){\circle*{8}}
\multiput(32.5,130)(0,-10){2}{\begin{picture}(5,10)
\put(2.5,10){\oval(5,5)[r]}
\put(2.5,5){\oval(5,5)[l]}\end{picture}}
\put(5,120){$b$}
\put(5,100){$\bar q$}
\put(5,110){$B$}
\put(65,120){$c$}
\put(65,100){$\bar q$}
\put(65,110){$D^{**}$}
\put(43,140){$W$}
\put(0,85){\small FIG. 1. Lowest order vertex function $\Gamma^{(1)}$
contributing to the current matrix element (18). }
\put(10,20){\line(1,0){50}}
\put(10,40){\line(1,0){50}}
\put(25,40){\circle*{8}}
\put(25,40){\thicklines \line(1,0){20}}
\multiput(25,40.5)(0,-0.1){10}{\thicklines \line(1,0){20}}
\put(25,39.5){\thicklines \line(1,0){20}}
\put(45,40){\circle*{2}}
\put(45,20){\circle*{2}}
\multiput(45,40)(0,-4){5}{\line(0,-1){2}}
\multiput(22.5,50)(0,-10){2}{\begin{picture}(5,10)
\put(2.5,10){\oval(5,5)[r]}
\put(2.5,5){\oval(5,5)[l]}\end{picture}}
\put(5,40){$b$}
\put(5,20){$\bar q$}
\put(5,30){$B$}
\put(65,40){$c$}
\put(65,20){$\bar q$}
\put(65,30){$D^{**}$}
\put(33,60){$W$}
\put(90,20){\line(1,0){50}}
\put(90,40){\line(1,0){50}}
\put(125,40){\circle*{8}}
\put(105,40){\thicklines \line(1,0){20}}
\multiput(105,40.5)(0,-0.1){10}{\thicklines \line(1,0){20}}
\put(105,39,5){\thicklines \line(1,0){20}}
\put(105,40){\circle*{2}}
\put(105,20){\circle*{2}}
\multiput(105,40)(0,-4){5}{\line(0,-1){2}}
\multiput(122.5,50)(0,-10){2}{\begin{picture}(5,10)
\put(2.5,10){\oval(5,5)[r]}
\put(2.5,5){\oval(5,5)[l]}\end{picture}}
\put(85,40){$b$}
\put(85,20){$\bar q$}
\put(85,30){$B$}
\put(145,40){$c$}
\put(145,20){$\bar q$}
\put(145,30){$D^{**}$}
\put(133,60){$W$}
\put(0,5){\small FIG. 2. Vertex function $\Gamma^{(2)}$
taking the quark interaction into account. Dashed lines correspond  }
\put(0,0) {\small to the effective potential
(\ref{qpot}). Bold lines denote the negative-energy part of the quark
propagator. }

\end{picture}

\end{figure}
\setcounter{figure}2

\begin{figure}
\begin{turn}{-90}
\epsfxsize=14 cm
\epsfbox{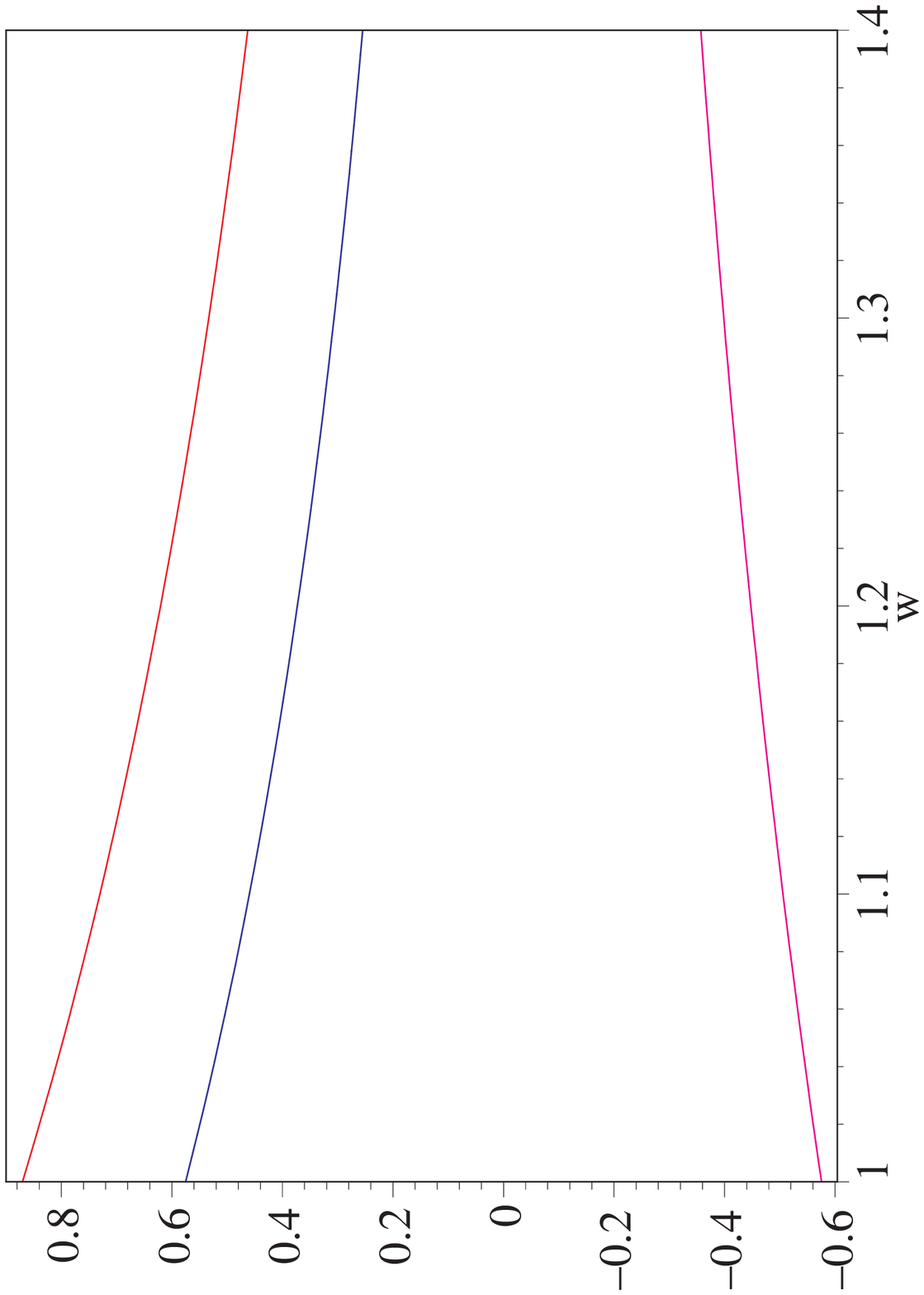}
\end{turn}

\caption{Isgur--Wise functions $\tau(w)$ (upper curve), $\tau_1(w)$
(middle curve) and $\tau_2(w)$ (lower curve) for the $B\to D_{1,2}e\nu$
decay.}
\end{figure}

\begin{figure}
\begin{turn}{-90}
\epsfxsize=14 cm
\epsfbox{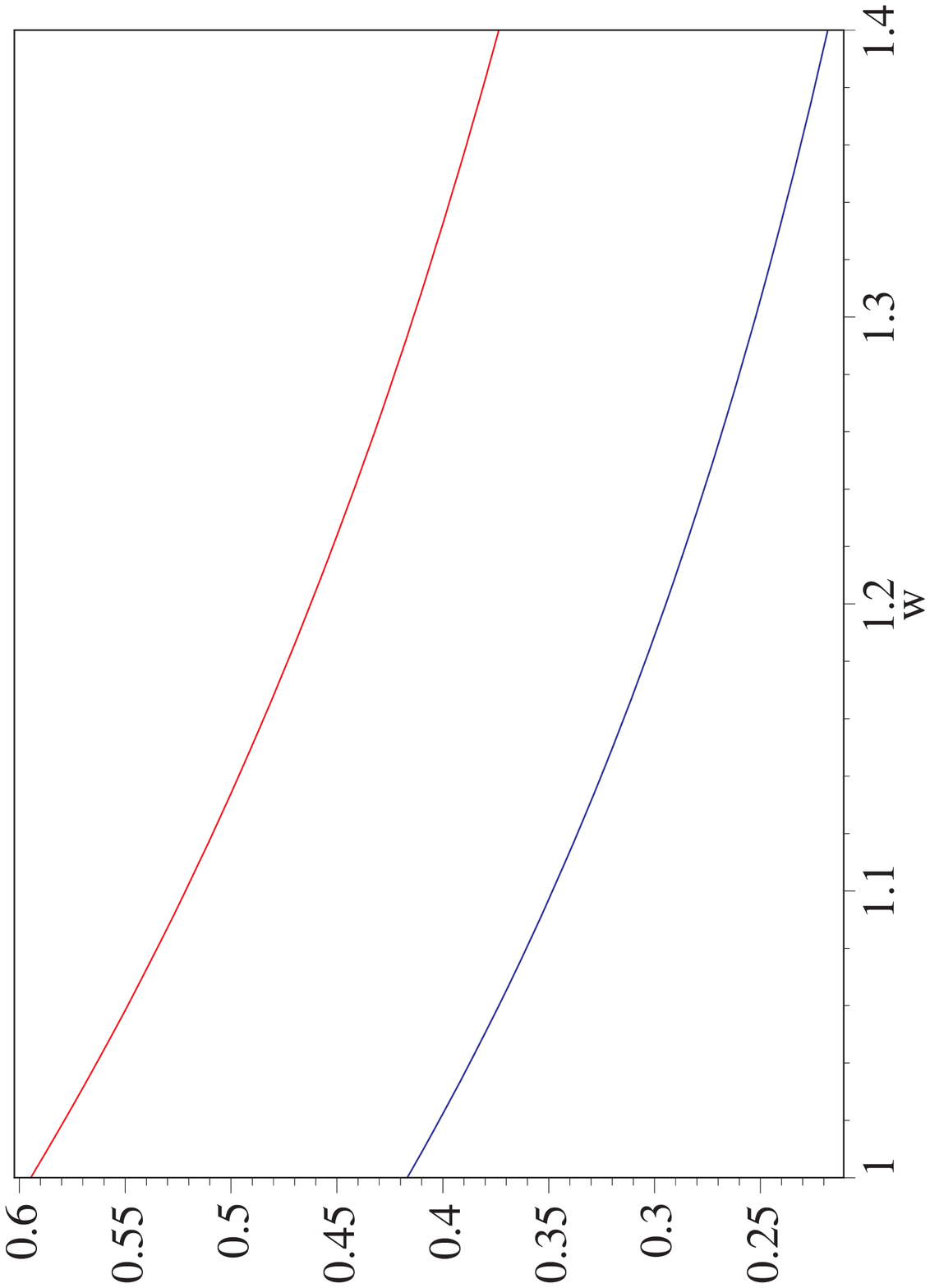}
\end{turn}

\caption{Isgur--Wise functions $\zeta(w)$ (upper curve) and
$\zeta_1(w)$ (lower curve) for the $B\to D^*_{0,1}e\nu$
decay.}

\end{figure}

\begin{figure}
\begin{turn}{-90}
\epsfxsize=14 cm
\epsfbox{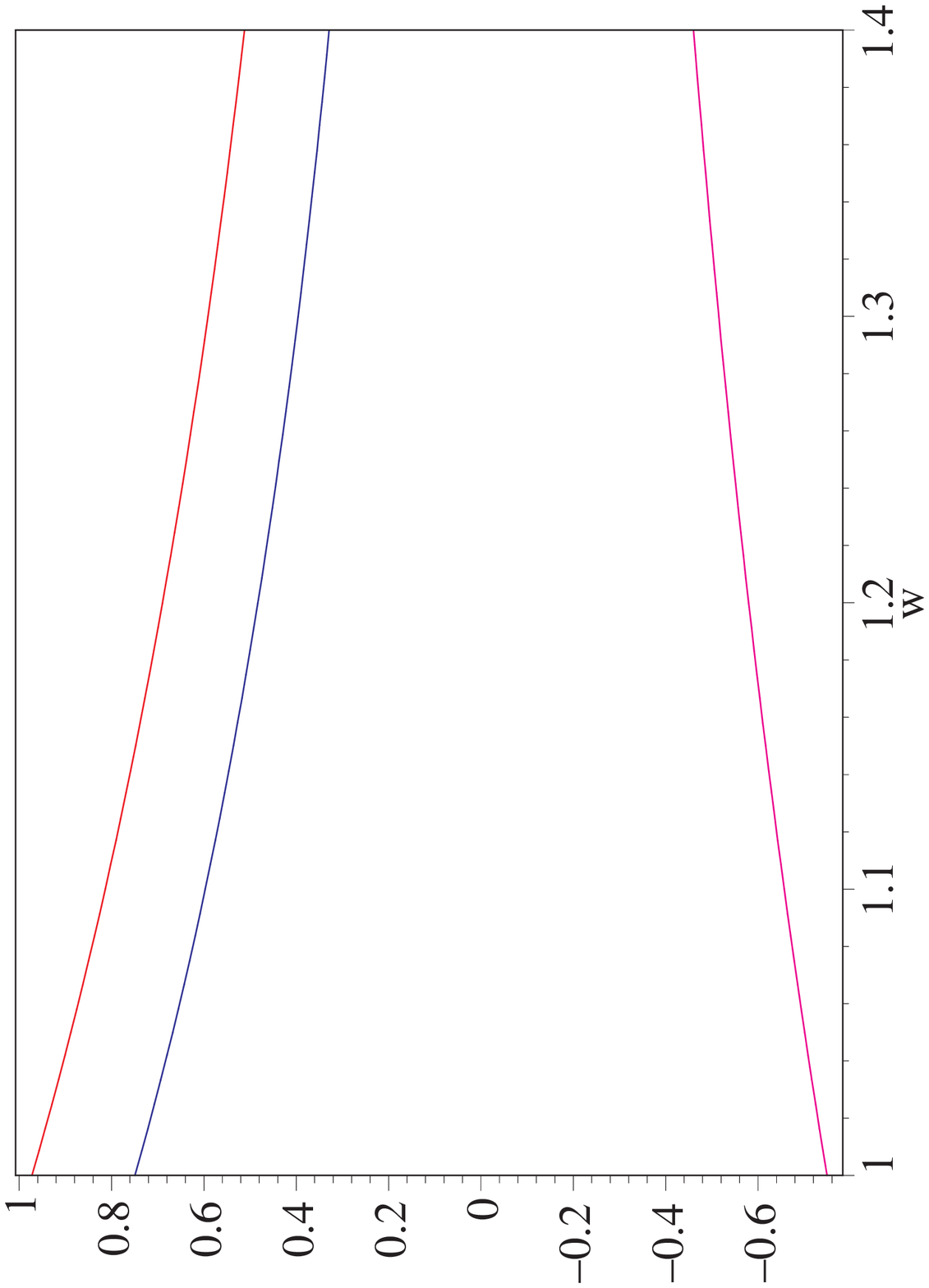}
\end{turn}

\caption{Isgur--Wise functions $\tau(w)$ (upper curve), $\tau_1(w)$
(middle curve) and $\tau_2(w)$ (lower curve) for the $B_s\to D_{s1,2}e\nu$
decay.}
\end{figure}

\begin{figure}
\begin{turn}{-90}
\epsfxsize=14 cm
\epsfbox{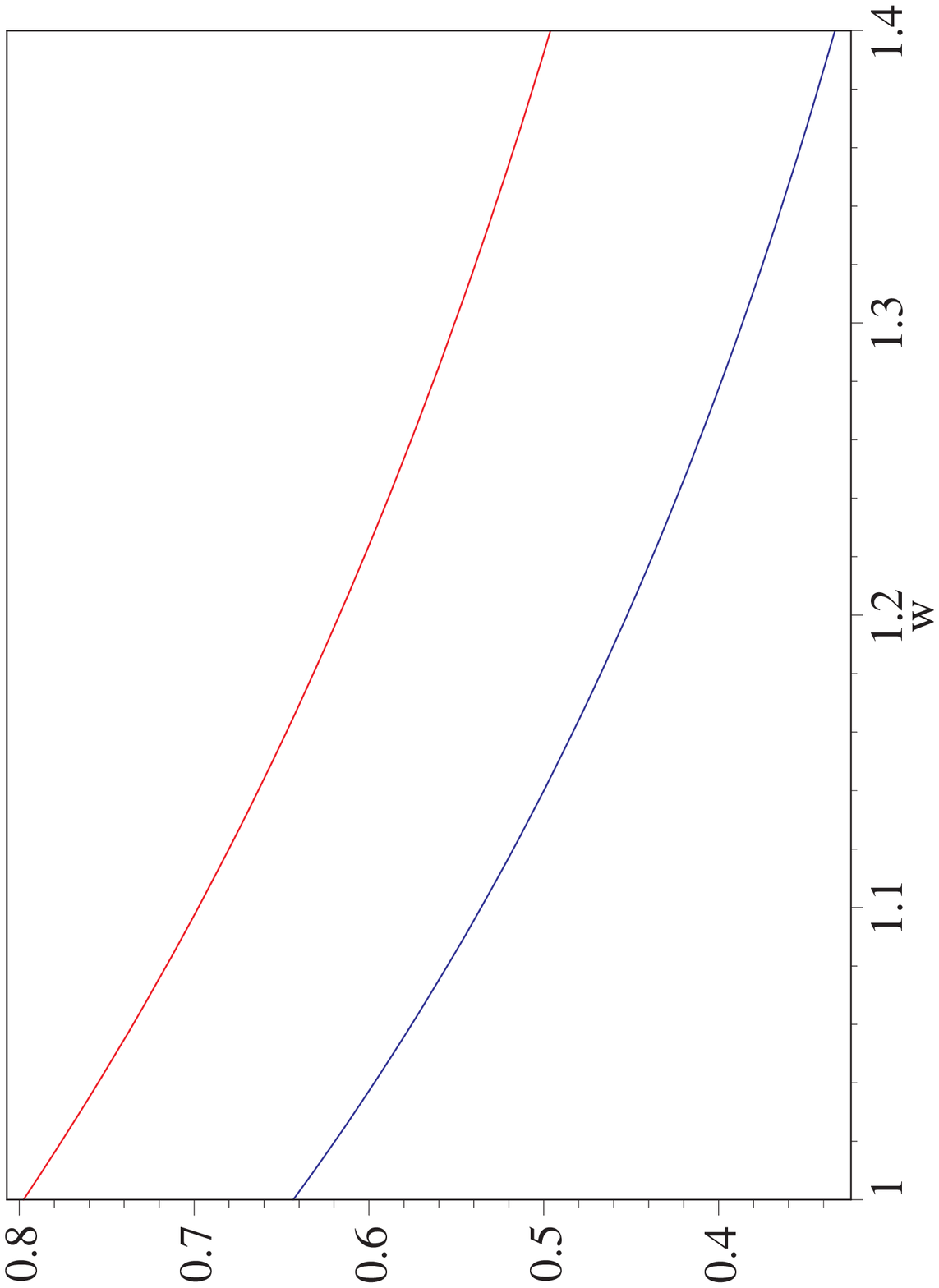}
\end{turn}

\caption{Isgur--Wise functions $\zeta(w)$ (upper curve) and
$\zeta_1(w)$ (lower curve) for the $B_s\to D^*_{s0,1}e\nu$
decay.}
\end{figure}

\twocolumn

\begin{figure}
\begin{flushleft}
\begin{turn}{-90}
\epsfxsize=8 cm
\epsfbox{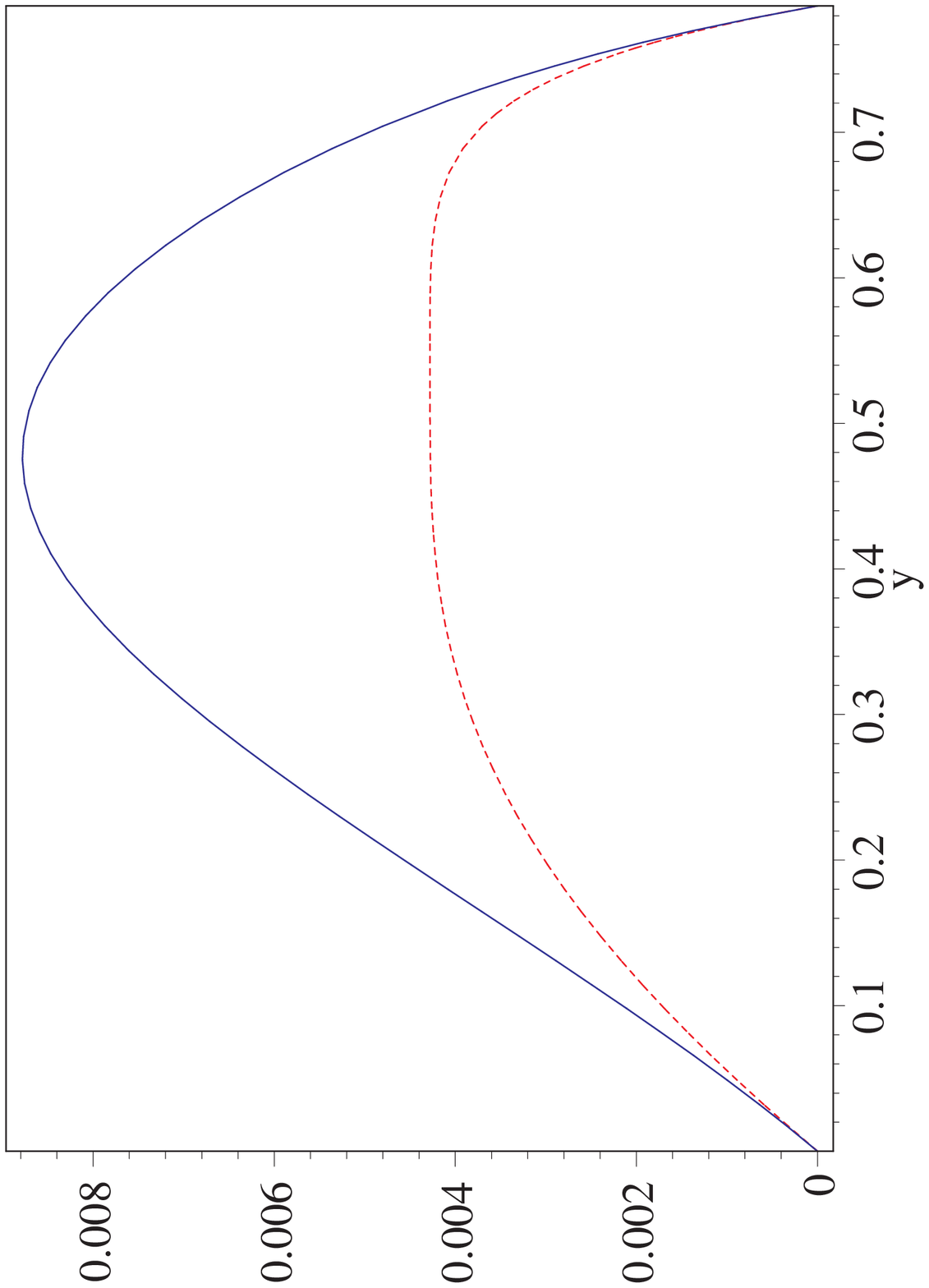}
\end{turn}
\centerline{(a) $B\to D^{*}_0 e\nu$}
\end{flushleft}
\end{figure}

\begin{figure}
\begin{flushleft}
{\begin{turn}{-90}
\epsfxsize=8 cm
\epsfbox{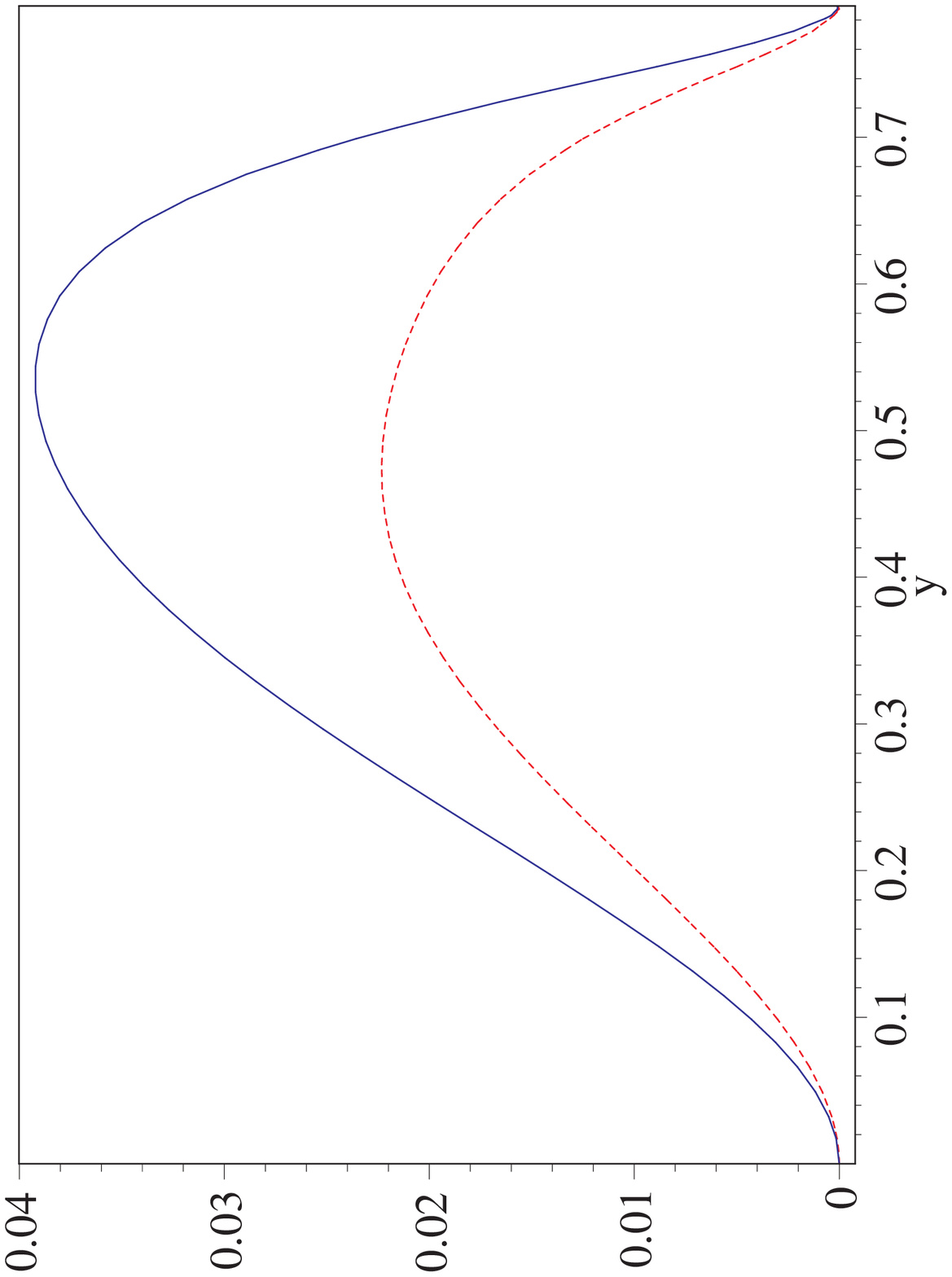}
\end{turn}}

\centerline{(c) $B\to D_1 e\nu$}
\end{flushleft}
\end{figure}

\widetext{
  \hbox to 17 cm{\hskip 0.5 cm\small FIG.~ 7. {Electron spectra
$\left({1}/{\Gamma_0}\right)\left(
{{\rm d}\Gamma}/{{\rm d
} y}\right)$ for the $B\to D^{**}e\nu$ decays.
  Dashed curves show the $m_Q\to\infty$}} 
\noindent \hbox to 10 cm {limit, solid curves include first
order $1/m_Q$ corrections.} }

\begin{figure}
\begin{flushleft}
\begin{turn}{-90}
\epsfxsize=8 cm
\epsfbox{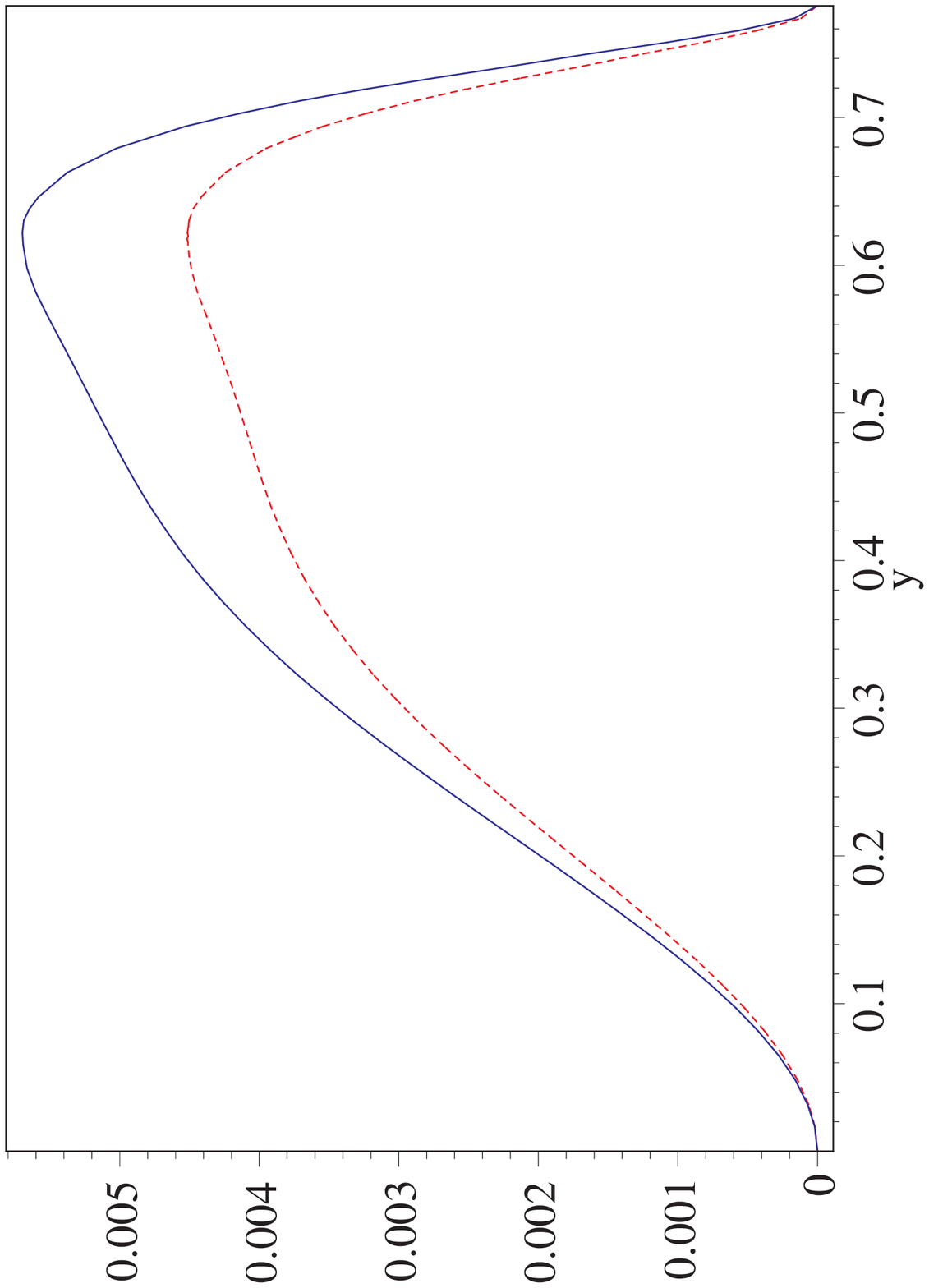}
\end{turn}


\centerline{(b) $B\to D^{*}_1 e\nu$}
\end{flushleft}
\end{figure}

  \begin{figure}
\begin{flushleft}
\begin{turn}{-90}
\epsfxsize=8 cm
\epsfbox{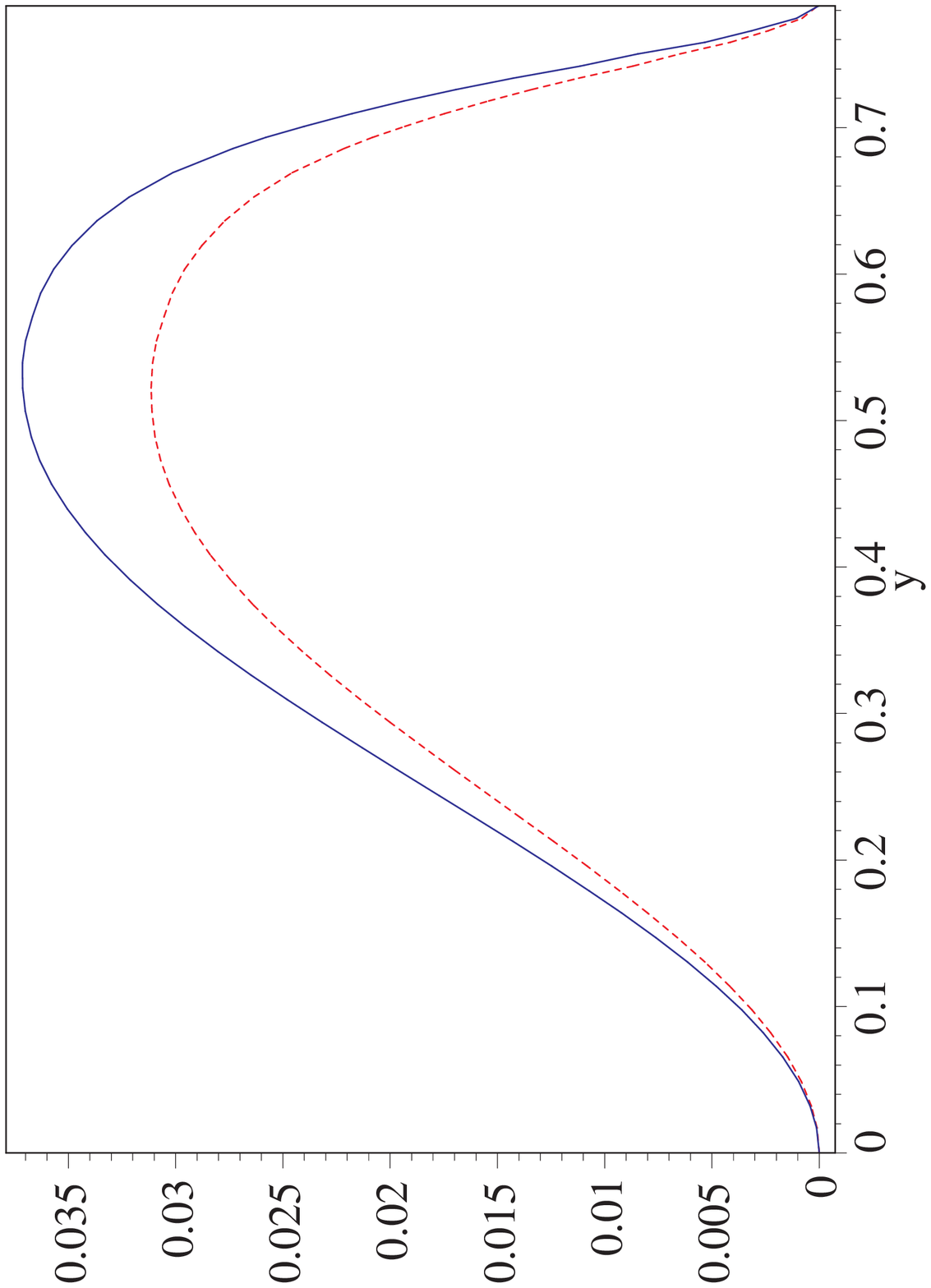}
\end{turn}

\centerline{(d) $B\to D^{*}_2 e\nu$}
\end{flushleft}
\end{figure}

\onecolumn

\end{document}